\documentclass[12pt, a4paper]{article}
\usepackage{a4wide,amssymb,amsmath,amscd}

\newcommand{\T}{{\mathbb T}}
\newcommand{\Z}{{\mathbb Z}}
\newcommand{\R}{{\mathbb R}}
\newcommand{\C}{{\mathbb C}}
\newtheorem{theorem}{Theorem}
\newtheorem{proposition}[theorem]{Proposition}

\begin{document}

\topmargin -2pt

\headheight 0pt

\topskip 0mm \addtolength{\baselineskip}{0.20\baselineskip}
\begin{flushright}
{\tt KIAS-P07020}
\end{flushright}


\begin{center}
{\Large \bf Quantum Thetas on Noncommutative ${\T}^d $  \\
 with General Embeddings } \\

\vspace{10mm}

{\sc Ee Chang-Young}\footnote{cylee@sejong.ac.kr}\\
{\it Department of Physics, Sejong University, Seoul 143-747, Korea}\\

{\it School of Physics, Korea Institute for Advanced Study,
Seoul 130-722, Korea}\\

\vspace{2mm}

and \\

\vspace{2mm}

{\sc Hoil Kim}\footnote{hikim@knu.ac.kr}\\

{\it Topology and Geometry Research Center, Kyungpook National University,\\
Taegu 702-701, Korea}\\

\vspace{10mm}

{\bf ABSTRACT} \\
\end{center}

\vspace{2mm}

\noindent
In this paper we construct quantum theta functions over
noncommutative ${\T}^d$ with general embeddings. Manin has
constructed quantum theta functions from the lattice embedding
into vector space $\times$ finite group. We extend Manin's
construction of quantum thetas to the case of general embedding of
vector space $\times$ lattice $\times$ torus. It turns out that
only for the vector space part of the embedding there exists the
holomorphic theta vector, while for the lattice part there does
not. Furthermore, the so-called quantum translations from
embedding into the lattice part become non-additive, while those
from the vector space part are additive.
\\



\thispagestyle{empty}

\newpage
\section{Introduction}

In the quantization of classical theta function, we encounter two
types of objects. One is the theta vector introduced by
Schwarz\cite{schwarz01}, which is a holomorphic element of a
projective module over unitary quantum torus. The other is the
quantum theta function introduced by
Manin\cite{manin1,manin1-tr,manin2,manin3}, which is an element of
the function ring of quantum torus itself. This is a natural
outcome if we consider the process of quantization, in which
commutative physical observables become operators acting on the
states. Namely, classically we have only one type of objects,
observables, and then after quantization we come up with two types
of objects, operators and states. This is exactly what happens
here. In the classical case, a set of specific values of
observables constitutes a state, and the classical theta function
is just like a state function. On the other hand, the quantum
theta functions and the theta vectors are oprators and state
vectors, respectively, in the quantum case.
Manin\cite{manin2,manin3} has shown that the Rieffel's algebra
valued inner(scalar) products \cite{rief88} of theta
vectors\cite{ds02} obtained from the lattice embedding of the type
$ {\R}^p (\times F)$ for quantum torus satisfy the property of
quantum theta function that he defined. Here, $d= 2p $ is the
dimension of the relevant quantum torus and $F$ is a finite group.
However, it was also shown in \cite{rief88} that there is another
type of lattice embedding for quantum torus, $ {\R}^p \times
{\Z}^q (\times F )$, where the dimension of the relevant quantum
torus is $d= 2p + q$. Manin has left the construction of the
quantum theta funciton for this case in question\cite{manin3}.

This type of non-zero $q$ embedding is intimately related to the
Morita equivalence over noncommutative tori \cite{rs98}. In
\cite{ek05}, we investigated the symmetry of quantum torus,
restricting ourselves to the symmetry of the algebra and its
module, which is not related to the Morita equivalence.
In that case, we only considered the embeddings with $q=0$.
However, to investigate the full symmetry of noncommutative tori
including the Morita equivalence, we need to understand the
behavior of modules from non-zero $q$ embeddings.

We have previously constructed the quantum theta function in the
latter type of embeddings that Manin has left in question
 in the case of noncommutative ${\T}^4 $ \cite{ek06}. This paper is the
extension of the work in \cite{ek06} to higher dimensional tori,
providing the general proof of the result of the ${\T}^4 $ case
extended to arbitrary ${\T}^d $ case.

We first try to find the theta vector in the non-zero $q$
embedding, and end up with a conclusion that holomorphic theta
vector does not exist in a general sense. Then we try to construct
the quantum theta function in this case.
Because still there is a possibility that
the Rieffel's scalar product with an element of non-holomorphic
(partially holomorphic only for the ${\R}^p$-part) module in the
second type of embedding satisfies the required property of the
quantum theta function. Thus we construct a quantum theta function
via Rieffel's scalar product with an element of the module in the
second type of embedding and find that it satisfies the
requirement of quantum theta function.
The organization of the paper is as follows. In the section 2, we
construct the modules with general embeddings for quantum tori. In
the section 3, we construct the quantum theta functions evaluating
the scalar products of the above obtained modules, and check the
required conditions for the quantum theta function. In the section
4, we conclude with discussion.




\section{Lattice embedding of quantum torus}\label{embedd}

We first review the embedding of quantum torus \cite{rief88} and a
canonical construction of the module with an embedding of the type
$ {\R}^p $, of which the four-torus case was done explicitly in
\cite{kl03}. Then we proceed to the case with an embedding of the
type  $ {\R}^p \times {\Z}^q $.

Recall that ${\T}^d_\theta$ is a deformed algebra of the algebra
of smooth functions on the torus ${\T}^d$ with the deformation
parameter $\theta$, which is a real $d\times d$ anti-symmetric
matrix. This algebra is generated by operators $U_1,\cdots,U_d$
obeying the following relations
\begin{align*}
U_jU_i=e^{2\pi i \theta_{ij}}U_iU_j \text{ \ and \ }
U_i^*U_i=U_iU_i^*=1, \text{ \ \ } i,j=1,\cdots,d.
\end{align*}
The above relations define the presentation of the involutive
algebra
$${\cal A}_\theta^d=
\{\sum a_{i_1\cdots i_d}U_1^{i_1}\cdots U_d^{i_d}\mid
a=(a_{i_1\cdots i_d})\in {\cal S}({\Z}^d)\}$$ where ${\cal
S}({\Z}^d)$ is the Schwartz space of sequences with rapid decay.

Every projective module over a smooth algebra ${\cal
A}^{d}_{\theta}$ can be represented by a direct sum of modules of
the form ${\cal S}({\R}^p\times{\Z}^q\times F)$, the linear space
of Schwartz functions on ${\R}^p\times{\Z}^q\times F$, where
$2p+q=d$ and $F$ is a finite abelian group.
 The module action is specified
by operators on ${\cal S}({\R}^p\times{\Z}^q\times F)$ and the
commutation relation of these operators should be matched with
that of elements in ${\cal A}^{d}_{\theta}$.

 Recall that there is the dual action of
the torus group ${\T}^d$ on ${\cal A}_\theta^d$ which gives a Lie
group homomorphism of ${\T}^d$ into the group of automorphisms of
${\cal A}_\theta^d$. Its infinitesimal form generates a
homomorphism of Lie algebra $ L$ of ${\T}^d$ into Lie algebra of
derivations of ${\cal A}_\theta^d$. Note that the Lie algebra $L$
is abelian and is isomorphic to ${\R}^d$. Let $\delta:L\rightarrow
{\rm {Der \ }}({\cal A}_\theta^d)$ be the homomorphism. For each
$X\in L$, $\delta(X):=\delta_X$ is a derivation i.e., for $u,v\in
{\cal A}_\theta^d$,
\begin{equation} \label{derv1}
\delta_X(uv)=\delta_X(u)v+u\delta_X(v).
\end{equation}
Derivations corresponding to the generators $\{e_1,\cdots,e_d\}$
of $L$ will be denoted by $\delta_1,\cdots,\delta_d$. For the
generators $U_i$'s of ${\T}_\theta^d$, it has the following
property
\begin{equation} \label{derv2}
\delta_i(U_j)=2\pi i\delta_{ij} U_j.
\end{equation}
Let $D$ be a lattice in ${\cal G}=M\times \widehat{M}$, where
$M={\R}^p\times{\Z}^q\times F$ and $\widehat{M}$ is its dual. Let
$\Phi$ be an embedding map such that $D$ is the image of ${\Z}^d$
under the map $\Phi$. This determines a projective module to be
denoted by $E$ \cite{rief88}. If $E$ is a projective ${\cal
A}_\theta^d$-module, a connection $\nabla$ on $E$ is a linear map
from $E$ to $E\otimes L^*$ such that for all $X\in L$,
\begin{align} \label{conn1}
\nabla_X(\xi u)=(\nabla_X\xi)u+\xi\delta_X(u),{\rm { \ \ \
}}\xi\in {E}, u\in {\cal A}_\theta^d.
\end{align}
It is easy to see that
\begin{align} \label{conn2}
[\nabla_i,U_j]=2\pi i\delta_{ij} U_j.
\end{align}

\noindent
In the Heisenberg representation the operators are
 defined by
\begin{equation}\label{pirep}
{\cal U}_{(m,\hat s)}f(r)=e^{2\pi i <r, \hat s>}f(r+m)
\end{equation}
for $(m,\hat s)\in\ D, \ r \in M .$

Now, we proceed to the construction of the module, first for the
embedding with the type $M= {\R}^p$, then with the type $M= {\R}^p
\times {\Z}^q $. Here we suppress the finite part for brevity.
%
%
%
%
We consider the embeddings of canonical forms in the present
section, and in the next section we will further consider the
generalization of the result from the canonical embeddings.

 For $M= {\R}^p$ with $2p=d$, we put the embedding map as
follows via proper rearrangement of the basis,
%
\begin{align}\label{phii}
\Phi_{\rm irr}=\begin{pmatrix}\Theta &0 \\
                0&   I \end{pmatrix} : = (x_{i,j}), \ \ {\rm for} \ \ i,j=1,
                \cdots, d,
\end{align}
where $\Theta$ and $I$ belong to ${\R}^p $ and ${{\R}^p}^*$,
respectively, and
 are given by $p\times p$ diagonal matrices of the type
\begin{align}\label{Theta}
\Theta = {\rm diag}(\theta_1, \cdots, \theta_p), \ \  I =
(\delta_{ij}), \ \ i,j = 1, \cdots, p.
\end{align}
Then using the expression (\ref{pirep}) for the Heisenberg
representation, we get
\begin{align}
& (U_j f)(s_1, \cdots, s_p)  : =  (U_{e_j}f)(\vec{s}),  \notag \\
    & \equiv  \exp(2\pi i \sum_{k=1}^p s_k x_{k+p,j} + \sum_{k=1}^p x_{k,j} x_{p+k,j} )f(\vec{s} +\vec{x}_j), \\
    & ~ ~~ {\rm for} \ ~ \ j=1,
\cdots, 2p  , \notag
\end{align}
where $\vec{s}=(s_1, \cdots, s_p), ~ \vec{x}_j=(x_{1,j}, \cdots,
x_{p,j})$ and $ \vec{s}, \vec{x}_j \in {\R}^p $.

\noindent This can be redisplayed as
\begin{align} \label{uops}
(U_j f)(\vec{s})&= f(\vec{s}  + \vec{\theta}) , \notag \\
(U_{j+p} f)(\vec{s})&=e^{2\pi i s_j } f(\vec{s}), \ \ {\rm for} \
\ j,k=1,
                \cdots, p,
\end{align}
where $\vec{\theta}=(\theta_1, \cdots, \theta_p)$. One can see
that they satisfy
\begin{align} \label{Ucomm}
U_j U_{j+p} & =e^{2\pi i \theta_j}U_j U_{j+p},
\end{align}
and otherwise $U_j U_k = U_k U_j $.
\\

%

 For the embedding of the type $ M= {\R}^p \times {\Z}^q $ where $2p+q= d$,
we put the embedding map of the canonical form as follows.
\begin{align}\label{embqnz}
\Phi_{\rm irr}=\begin{pmatrix}\Theta & 0 & 0 \\
                0& I & 0 \\ 0 & 0 & Q \\ 0 & 0 & \Delta  \end{pmatrix} : = (x_{i,j}),
                ~ ~ i=1,..,2p+2q, ~ ~ j=1,..,2p+q,
\end{align}
where $\Theta $ and $I$ are the same as before that belong to $
{\R}^p $ and $ {{\R}^p}^*$, respectively, and $Q$ and $\Delta$ are
$q\times q$ matrices that belong to ${\Z}^q$ and $ T^q$,
respectively.
Then, the operators $U_j$ acting on the space $E :={\mathcal
S}({\mathbb R}^p \times {\Z}^q ) $ can be defined via Heisenberg
representation (\ref{pirep}), and we get
\begin{align}
& (U_j f)(s_1,\cdots, s_p, n_1, \cdots n_q)
   : =
(U_{e_j}f)(\vec{s},\vec{n}), \notag  \\
    & \equiv  e^{2\pi i (
\sum_{k=1}^p s_k x_{p+k,j} + \sum_{l=1}^q n_l x_{2p+q+l,j})+ \pi i
(\sum_{k=1}^p x_{k,j} x_{p+k,j} + \sum_{l=1}^q x_{2p+l,j}
x_{2p+q+l,j}) }  f(\vec{s} + \vec{x}_{1j}, \vec{n} + \vec{x}_{2j}), \notag \\
                 & ~ ~~ {\rm for} \  \ j=1, \cdots, 2p+q  ,
\label{embed}
\end{align}
where $ ~ \vec{x}_{1j}= (x_{1,j}, \cdots, x_{p,j})$ and $~
\vec{x}_{2j}= (x_{2p+1,j}, \cdots, x_{2p+q,j})$  belonging to
${\mathbb R}^p , {\Z}^q$, respectively.
\\

%
\section{Quantum thetas}\label{qtheta}

In this section, we first try to construct the theta vector by
defining the connection with a complex structure for the embedding
of the type  $ {\R}^p \times {\Z}^q $. Then, we construct the
quantum theta function following the Manin's construction.

%
\subsection*{3.1 Theta vectors}\label{thevec}




In the previous section, connections on a projective ${\cal
A}_\theta^d$-module satisfies the condition (\ref{conn2}) and it
can be written as
\begin{align}
\label{concr}
 U_j \nabla_i = \nabla_i U_j - 2 \pi i \delta_{ij} U_j , ~ ~ {\rm for} ~ ~ i,j = 1, \cdots, 2p+q.
\end{align}
%




\noindent

\begin{proposition}\rm{\bf{(Rieffel)}} \it{ The relation (\ref{concr}) is satisfied
with the connection $\nabla_j$ such that
\begin{align}
(\nabla_j f)(\vec{s},\vec{n})& = - 2\pi i (\sum_{k=1}^p  B_{j,k}
s_k f(\vec{s},\vec{n}) + \sum_{l=1}^q
 B_{j,2p+l}n_l f(\vec{s},\vec{n}) ) \notag \\
& ~~  + \sum_{k=1}^p B_{j, p+k} \frac{\partial f}{\partial s_k}
(\vec{s},\vec{n}), ~~ \text{for} ~~ j=1,..,2p+q,
\label{conn-rl}
\end{align}
where $\vec{s}=(s_1, \cdots, s_p), ~ \vec{n}=(n_1, \cdots, n_q)$,
and the constants $B_{j,k} \in \R$ satisfy the following
condition,

%
%
\begin{equation}
\label{bmatrix1}
\sum_{k=1}^p (B_{i,k}x_{k,j}+ B_{i,p+k}x_{p+k,j})
+ \sum_{l=1}^q B_{i,2p+l}x_{2p+l,j} = \delta_{ij} , ~~~ i,j
=1,..,2p+q.
\end{equation}
}
\end{proposition}

\vspace{1cm}

The condition (\ref{bmatrix1}) says that the matrix $B$ is the
inverse matrix of $\tilde{X}$ where $\tilde{X}_{ij}=(x_{i,j})$ for
$i,j=1, \cdots, 2p+q$. Namely, the inverse matrix of the upper
$(2p + q) \times (2p+q)$ part of the matrix $(x_{i,j})$ is the
matrix $B$:
\begin{equation}
\label{bmatrix2}
B = {\tilde{X}}^{-1} , ~~ {\rm and} ~~ \tilde{X} = \begin{pmatrix}\Theta & 0 & 0 \\
                0& I & 0 \\ 0 & 0 & Q  \end{pmatrix} ,
\end{equation}
where  $\Theta, I, Q$ are given for the canonical form in
(\ref{embqnz}).




We say that a noncommutative torus is equipped with complex
structure if the Lie algebra $L$ mentioned in the section 2 is
equipped with such a structure.
A complex structure on $L$ can be
considered as a decomposition of complexification $L \oplus
 iL$ of $L$ in a direct sum of two complex conjugate subspace
 $L^{1,0}$ and $L^{0,1}$.
 We denote by $\bar{\delta}_1,\dots ,
 \bar{\delta}_{d/2}$, a basis in $L^{0,1}$. One can express
 $\bar{\delta}_{\alpha}, ~ \alpha = 1, \dots, d/2$ in terms of
 $\delta_{\beta}, ~ \beta = 1, \dots , d $ which appeared in the
 section 2 as $\bar{\delta}_{\alpha}= h^{\beta}_{\alpha}
 \delta_{\beta}$, where $h^{\beta}_{\alpha}$ is a complex
 $\frac{d}{2} \times d$ matrix.
 A complex structure on a ${\cal
 A}_\theta^d$-module $E$ can be defined as a collection of
 ${\C}$-linear operators $\overline{\nabla}_1, \dots,
 \overline{\nabla}_{d/2}$ on $E$ satisfying
\begin{align} \label{com-conn}
\overline{\nabla}_{\alpha}(a f)=  a \overline{\nabla}_{\alpha}(f)
 + \bar{\delta}_{\alpha} (a)  f ,  \ \ \
 a \in {\cal A}_\theta^d, ~  f \in E .
\end{align}
A vector $ f \in E$  is called holomorphic if
\begin{equation}
\label{hol-vec}
\overline{\nabla}_{\alpha} f = 0, ~~ \alpha = 1,
\dots, d/2 .
\end{equation}

\noindent
Now, we assume that there exists a complex structure $ T
$ such that
\begin{equation}
\label{cmplex-str}
\begin{pmatrix} \overline{\nabla}_1  \\
           \vdots \\
\overline{\nabla}_{d/2}
             \end{pmatrix} =
            \begin{pmatrix}
          T , ~ I
            \end{pmatrix}
\begin{pmatrix} \nabla_1  \\ \vdots  \\
          \nabla_d \end{pmatrix}
\end{equation}
where $T$ is a $\frac{d}{2} \times \frac{d}{2}$ complex matrix and
$I$ is a $\frac{d}{2} \times \frac{d}{2}$ unit matrix. In the
canonical embedding (\ref{embqnz}),
 the connection $\nabla_{\beta}$ is given by
(\ref{conn-rl}) and (\ref{bmatrix2})
\begin{equation}
\label{conn}
\begin{pmatrix} \nabla_1  \\
           \vdots \\
\nabla_d
             \end{pmatrix} =
            \begin{pmatrix}
          \Theta^{-1} & 0 & 0 \\ 0 & I & 0 \\ 0 & 0 & Q^{-1}
            \end{pmatrix}
\begin{pmatrix} -2 \pi i s_1  \\ \vdots  \\ -2 \pi i s_p \\
 \frac{\partial}{\partial s_1} \\ \vdots \\ \frac{\partial}{\partial
 s_p} \\  -2 \pi i n_1  \\ \vdots  \\ -2 \pi i n_q
           \end{pmatrix} .
\end{equation}
If there exists a holomorphic vector $f(\vec{s},\vec{n})$, then
the following equation should be satisfied:
\begin{equation}
\label{hol-vect1}
\begin{pmatrix} \overline{\nabla}_1  \\
           \vdots \\
\overline{\nabla}_{d/2}
             \end{pmatrix} f = 0 .
\end{equation}
The above can be written as
\begin{equation}
\label{hol-vect2}
  \begin{pmatrix}
          T , ~ I
            \end{pmatrix}
            \begin{pmatrix}
          \Theta^{-1} & 0 & 0 \\ 0 & I & 0 \\ 0 & 0 & Q^{-1}
            \end{pmatrix}
\begin{pmatrix} -2 \pi i s_1  \\ \vdots  \\ -2 \pi i s_p \\
 \frac{\partial}{\partial s_1} \\ \vdots \\ \frac{\partial}{\partial
 s_p} \\  -2 \pi i n_1  \\ \vdots  \\ -2 \pi i n_q
           \end{pmatrix} f =0 .
\end{equation}
 To check the existence condition for the holomorphic vector, we
let
\begin{equation}
\label{holcd}
  \begin{pmatrix}
          T , ~ I
            \end{pmatrix}
            \begin{pmatrix}
          \Theta^{-1} & 0 & 0 \\ 0 & I & 0 \\ 0 & 0 & Q^{-1}
            \end{pmatrix} :=
\begin{pmatrix}  A , ~ C , ~ F  \end{pmatrix}
\end{equation}
where $A$ and $C$ are $(p+\frac{q}{2})\times p $ matrices and $F$
is a $(p+\frac{q}{2})\times q $ matrix. Then the required
condition for $f$ is
\begin{align}
  2\pi i \sum_{k=1}^p ( A_{ik}
s_k  + F_{il} n_l) f   =  \sum_{k=1}^p C_{ik} \frac{\partial
f}{\partial s_k},  ~~~ {\rm for} ~ ~ i=1,\dots, \frac{d}{2} =p+
\frac{q}{2} . \label{cons-cd1}
\end{align}
The only possible function is of the form
\begin{equation}
\label{holvect} f(\vec{s}, \vec{n}) = \exp [2 \pi i (\frac{1}{2}
\sum_{j,k=1}^p s_j \Omega_{jk} s_k + \sum_{k=1}^p \sum_{l=1}^q
G_{lk}n_l s_k) ]
\end{equation}
where  $\Omega^t = \Omega$. Then the condition (\ref{cons-cd1})
becomes
\begin{align}
\sum_{k=1}^p C_{ik} \Omega_{kj} & = A_{ij}, ~~ 1 \leq i \leq
p+\frac{q}{2}, ~ 1 \leq j \leq p ,
 \notag \\
 \sum_{k=1}^p C_{ik} G_{lk} & = F_{il},  ~~ 1 \leq i \leq
p+\frac{q}{2}, ~ 1 \leq l \leq q .
   \label{cons-cd2}
\end{align}
In other words,
\begin{equation}
\label{cons-cd3}
C \Omega = A ~~
{\rm and} ~~ CG^t = F .
\end{equation}
Combining these two conditions and from (\ref{holcd}), we obtain
the following relation.
\begin{equation}
\label{cons-cd} C (\Omega , ~ I , ~ G^t )  = (A , ~ C , ~ F) = (T
, ~ I)
 \begin{pmatrix}
          \Theta^{-1} & 0 & 0 \\ 0 & I & 0 \\ 0 & 0 & Q^{-1}
            \end{pmatrix} .
\end{equation}
%
\\

\begin{proposition}
We consider the existence of the holomorphic vector in the
canonical embeddings in three different cases.
\\ (i) For $p\neq 0, ~ q=0 $, there is the unique holomorphic
vector with $\Omega= T \Theta ^{-1}$ which is symmetric and whose
imaginary part is positive definite.
\\ (ii) For $p \neq 0, ~ q \neq 0 $, the holomorphic vector does not exist.
\\ (iii) For $ p=0, ~ q\neq 0$, the only possible one is the delta function at the
origin.
\end{proposition}
\textbf{Proof.}
 In the case (i), the consistency relation
(\ref{cons-cd}) is reduced to
\begin{equation}
\label{cons-qzro} C (\Omega , ~ I )  = (A , ~ C) = (T , ~ I)
 \begin{pmatrix}
          \Theta^{-1} & 0  \\ 0 & I
            \end{pmatrix} = (T \Theta^{-1} , ~ I ) .
\end{equation}
Thus one can see immediately that $ C=I $ and $ \Omega = T
\Theta^{-1} $.
Since $\Omega$ is symmetric by construction, so is $T \Theta^{-1}
$, and this is the necessary condition for the existence of
holomorphic theta vector. Here, in order $f$ to be a Schwartz
function, the imaginary part of $ T \Theta^{-1} $ should be
positive.
\\
In the case (ii), the consistency relation (\ref{cons-cd}) is
\begin{equation}
\label{cons-pqnzro} C (\Omega , ~ I , ~ G^t )  = (T , ~ I)
 \begin{pmatrix}
          \Theta^{-1} & 0 & 0 \\ 0 & I & 0 \\ 0 & 0 & Q^{-1}
          \end{pmatrix} .
\end{equation}
The above relation can be understood as linear maps from
${\C}^{2p+q} \rightarrow {\C}^p \rightarrow {\C}^{p+\frac{q}{2}}$
for the left and from ${\C}^{2p+q} \rightarrow {\C}^{2p+q}
\rightarrow {\C}^{p+\frac{q}{2}}$ for the right.
The right linear map is surjective since both $(T , ~ I)$ and
$\begin{pmatrix}
          \Theta^{-1} & 0 & 0 \\ 0 & I & 0 \\ 0 & 0 & Q^{-1}
\end{pmatrix} $ are of full rank, while the left linear map cannot
be surjective  since it is maximally of rank $p$ which is strictly
smaller that $p+\frac{q}{2}$.
\\
In the case (iii), the consistency relation (\ref{cons-cd})
becomes
\begin{equation}
\label{cons-pzro} (T , ~ I)
 \begin{pmatrix}
           Q^{-1}
          \end{pmatrix}
           \begin{pmatrix}  -2 \pi i n_1  \\ \vdots  \\ -2 \pi i n_q
           \end{pmatrix} f =0 .
\end{equation}
If one can let $ (T , ~ I ) (Q^{-1}) = F$ as defined in
(\ref{holcd}), where $T$ and $I$ are $\frac{q}{2} \times
\frac{q}{2}$ matrices and $Q$ is $q \times q$ matrices, then the
above condition can be written as
\begin{align}
( \sum_{l=1}^q  F_{il} n_l) f (\vec{n})   = 0 ,  ~~~ {\rm for} ~ ~
i=1,\dots, \frac{q}{2} .
\label{holcd-pzro}
\end{align}
If $f$ should be a nontrivial solution, then $\sum_{l=1}^q  F_{il}
n_l=0$ for all $i=1,\dots, \frac{q}{2}$. Since $F_{il} \in {\C}$,
$(0, \cdots, 0)$ is the only solution for $\vec{n}$. Namely, $f$
can be nonzero only for $\vec{n}=(0, \cdots, 0)$, i.e., $f$ is a
delta function at the origin. And (\ref{holcd-pzro}), which is a
re-phrasal of (\ref{hol-vect1}), tells us that $f$ has
non-vanishing solution only when the connection vanishes. In
effect, one can say that the holomorphic vector does not exist in
this case, either.
\\



Now we consider the changes of the above result in the general
set-up. First consider the construction of the module from
embeddings of the type $M= {\R}^p \times {\Z}^q $ where $2p+q= d$.
Here again, we suppress the finite part for brevity.
%
Let the embedding map  be
\begin{align}\label{embqnz2}
\Phi : = (x_{i,j}), ~ ~ i=1,..,2p+2q, ~ ~ j=1,..,2p+q.
\end{align}
%
The operators $U_j$ acting on the space $E :={\mathcal S}({\mathbb
R}^p \times {\Z}^q ) $ can be defined via Heisenberg
representation, and are given by the equation (\ref{embed}) for
more general values of $x_{i,j}$ given by the above embedding.

For the theta vectors, equation (\ref{bmatrix1}) tells us that the
matrix $B$ is the inverse matrix of $\tilde{X}$ where
$\tilde{X}_{ij}=(x_{i,j})$ for $i,j=1, \cdots, 2p+q$. Namely, the
matrix $\tilde{X}$ is the upper $(2p + q) \times (2p+q)$ square
part of the matrix $\Phi $ and $B$ is its inverse matrix:
\begin{equation}
 \label{bmatrixg}
 B = {\tilde{X}}^{-1} .
\end{equation}

For a general complex structure, equation (\ref{cmplex-str}) can
be written as
\begin{equation}
\label{cmplex-str2}
\begin{pmatrix} \overline{\nabla}_1  \\
           \vdots \\
\overline{\nabla}_{d/2}
             \end{pmatrix} =
            \begin{pmatrix}
          T_1 , ~ T_2
            \end{pmatrix}
\begin{pmatrix} \nabla_1  \\ \vdots  \\
          \nabla_d \end{pmatrix}
\end{equation}
where $T_1$ and $T_2$ are $\frac{d}{2} \times \frac{d}{2}$ complex
matrices with $d$ given by $2p+q$. And the connection
$\nabla_{\beta}$ in (\ref{conn-rl}) becomes
\begin{equation}
\label{conn2}
\begin{pmatrix} \nabla_1  \\
           \vdots \\
\nabla_d
             \end{pmatrix} =
            \begin{pmatrix}
          B
            \end{pmatrix}
\begin{pmatrix} -2 \pi i s_1  \\ \vdots  \\ -2 \pi i s_p \\
 \frac{\partial}{\partial s_1} \\ \vdots \\ \frac{\partial}{\partial
 s_p} \\  -2 \pi i n_1  \\ \vdots  \\ -2 \pi i n_q
           \end{pmatrix}
\end{equation}
where $B$ is a $(2p+q) \times (2p+q)$ matrix defined by
(\ref{bmatrixg}). Now, the condition for holomorphic vector
(\ref{hol-vect1}) becomes
\begin{equation}
\label{hol-vectg}
  \begin{pmatrix}
          T_1 , ~ T_2
            \end{pmatrix}
            \begin{pmatrix}
          B
            \end{pmatrix}
\begin{pmatrix} -2 \pi i s_1  \\ \vdots  \\ -2 \pi i s_p \\
 \frac{\partial}{\partial s_1} \\ \vdots \\ \frac{\partial}{\partial
 s_p} \\  -2 \pi i n_1  \\ \vdots  \\ -2 \pi i n_q
           \end{pmatrix} f =0 .
\end{equation}
 To check the existence condition for the holomorphic vector
 we let
\begin{equation}
\label{holcd2}
  \begin{pmatrix}
          T_1 , ~ T_2
            \end{pmatrix}
            \begin{pmatrix}
          B
            \end{pmatrix} :=
\begin{pmatrix}  A , ~ C , ~ F  \end{pmatrix}
\end{equation}
where $A$ and $C$ are $(p+\frac{q}{2})\times p $ matrices and $F$
is a $(p+\frac{q}{2})\times q $ matrix. Then the holomorphic
condition for $f$ given by (\ref{holvect}) is the same as
 in
(\ref{cons-cd3}), and in the above notation, we can write the
following relation.
\begin{equation}
\label{cons-cdg} C (\Omega , ~ I , ~ G^t )  = (A , ~ C , ~ F) =
(T_1 , ~ T_2)
 \begin{pmatrix}
    B            \end{pmatrix} .
\end{equation}
\\
\begin{theorem}
 The existence of holomorphic vectors in the general embeddings is as follows:
\\(i) For $p\neq 0, ~ q=0 $, the unique solution is given by
\[
\Omega  = (T_1 B_{12} + T_2 B_{22})^{-1} (T_1 B_{11} + T_2
B_{21}),\]
where
\begin{equation}
B=
 \begin{pmatrix}
          B_{11} & B_{12}  \\ B_{21} & B_{22}
            \end{pmatrix},
~~B_{i,j}~\text{is}~p \times p~\text{matrix,}
\end{equation}
with three following conditions; (1) There should exist an inverse
of the matrix $(T_1 B_{12} + T_2 B_{22})$, (2) the matrix $(T_1
B_{12} + T_2 B_{22})^{-1} (T_1 B_{11} + T_2 B_{21})$ should be
symmetric,and (3) $ {\rm Im}((T_1 B_{12} + T_2 B_{22})^{-1} (T_1
B_{11} + T_2 B_{21}))
> 0$
\\(ii) For $p \neq 0, ~ q \neq 0 $, there does not exist holomorphic vector.
\\ (iii) For $ p=0,~ q\neq 0$, the only possible solution is the delta function at the origin.\\
\end{theorem}
\textbf{Proof.} In the case (i), the consistency relation
(\ref{cons-cdg}) is reduced to
\begin{equation}
\label{cons-qzro} C (\Omega , ~ I )  = (A , ~ C) = (T_1 , ~ T_2)
 \begin{pmatrix}
          B_{11} & B_{12}  \\ B_{21} & B_{22}
            \end{pmatrix} = (T_1 B_{11} + T_2 B_{21} , T_1 B_{12} + T_2 B_{22} )
\end{equation}
where we write the matrix $B$ in $2 \times 2$ block form with each
block being  a $p \times p$ matrix. Here, $\Omega$ is given by
\[ \Omega = (T_1 B_{12} + T_2 B_{22})^{-1} (T_1 B_{11} + T_2
B_{21}).
\]
In order to have a holomorphic theta vector the following
conditions should be satisfied: (1) There should exist an inverse
of the matrix $(T_1 B_{12} + T_2 B_{22})$, (2) the matrix $(T_1
B_{12} + T_2 B_{22})^{-1} (T_1 B_{11} + T_2 B_{21})$ should be
symmetric, since $\Omega$ is symmetric by construction, and (3) $
{\rm Im}((T_1 B_{12} + T_2 B_{22})^{-1} (T_1 B_{11} + T_2 B_{21}))
> 0 $ in order $f$ to be a Schwartz function.
\\
In the case (ii), the consistency relation (\ref{cons-pqnzro})
becomes
\begin{equation}
\label{cons-pqnzrog} C (\Omega , ~ I , ~ G^t )  = (T_1 , ~ T_2)
          B .
\end{equation}
The above relation can be understood as before in terms of linear
maps from ${\C}^{2p+q} \rightarrow {\C}^p \rightarrow
{\C}^{p+\frac{q}{2}}$ for the left, and from ${\C}^{2p+q}
\rightarrow {\C}^{2p+q} \rightarrow {\C}^{p+\frac{q}{2}}$ for the
right.
The right linear map is surjective since both $(T_1 , ~ T_2)$ and
$ B $ are of full rank, while the left linear map cannot be
surjective  since it is maximally of rank $p$ which is strictly
smaller that $p+\frac{q}{2}$ as before.
\\
In the case (iii), the relation (\ref{cons-pzro}) becomes
\begin{equation}
\label{cons-pzrog} (T_1 , ~ T_2)
 \begin{pmatrix}
           B
          \end{pmatrix}
           \begin{pmatrix}  -2 \pi i n_1  \\ \vdots  \\ -2 \pi i n_q
           \end{pmatrix} f =0 .
\end{equation}
If one can let $ (T_1 , ~ T_2)B = F$ as defined in (\ref{holcd2}),
where $T_1$ and $T_2$ are $\frac{q}{2} \times \frac{q}{2}$
matrices and $B$ is $q \times q$ matrices, then the above
condition can be written as
\begin{align}
( \sum_{l=1}^q  F_{il} n_l) f (\vec{n})   = 0 ,  ~~~ {\rm for} ~ ~
i=1,\dots, \frac{q}{2} . \label{holcd-pzrog}
\end{align}
In the same vein, should $f$ be a nontrivial solution, then
$\sum_{l=1}^q F_{il} n_l=0$ for all $i=1,\dots, \frac{q}{2}$ as
before. Thus $f$ can be nonzero only for $\vec{n}=(0, \cdots, 0)$,
and (\ref{holcd-pzrog}),  a re-phrasal of (\ref{hol-vect1}), tells
us that $f$ can be a non-vanishing solution only when the connection
vanishes. Therefore holomorphic vector does not exist in this case.
$\Box$\\



The above analysis shows that one cannot have a holomorphic vector
over totally complexified ${\T}_\theta^d $ in the embedding of $
M={\R}^p \times {\Z}^q $ with nonzero $p$ and $q$. This can be
remedied by giving a complex structure only over the continuous
part of the embedding space, i.e., by giving a complex structure
to the connection components over $ {\R}^p \times {\R}^{p*} $.
Now, we implement this as follows.
\begin{align}
\label{holconng}
\begin{pmatrix}
\overline{\nabla}_1  \\
           \vdots \\
\overline{\nabla}_p
             \end{pmatrix} & =
            \begin{pmatrix}
          T_1  , ~ T_2
            \end{pmatrix}
\begin{pmatrix} \nabla_1  \\ \vdots  \\
          \nabla_{2p} \end{pmatrix} , \notag \\
\overline{\nabla}_{p+1} & =  \nabla_{2p+1} , \\
   & ~~  \vdots  \notag \\
\overline{\nabla}_{p+q} & = \nabla_{2p+q} , \notag
\end{align}
where $ T_1$ and $T_2$ are $p \times p$ complex matrices and give
the complex structure over $ {\R}^p \times {\R}^{p*} $. Then, the
holomorphic vector over this part satisfies
%
%
\begin{align}
\label{holthetap}
\begin{pmatrix}
 \overline{\nabla}_1  \\
           \vdots \\
\overline{\nabla}_p
             \end{pmatrix}   f (\vec{s}, \vec{n})  = 0 ,
\end{align}
whose solution is given by
\[
 f(\vec{s}, \vec{n}) = \exp ( \pi i  \sum_{j,k=1}^p s_j
\Omega_{jk} s_k ) g (\vec{n}).
\]
Since $f$ belongs to $ {\mathcal S}({\mathbb R}^p) \otimes
{\mathcal S} ( {\Z}^q ) $,  $g (\vec{n})$  belongs to ${\mathcal
S} ( {\Z}^q )$ and has to be a Schwartz function. Here, we choose
a simple Schwartz function for $g(\vec{n})$, and write the
function $ f (\vec{s}, \vec{n})$ as
\begin{equation}
\label{thevec2}
  f(\vec{s}, \vec{n}) = \exp [ \pi i \sum_{j,k=1}^p s_j
\Omega_{jk} s_k  - \frac{\pi}{2}
\sum_{i=1}^{\frac{q}{2}}(n_i^2+n_{\frac{q}{2}+i}^2)] ,
\end{equation}
where ${\rm Im}\Omega > 0$.



\subsection*{3.2 Quantum theta functions}\label{qthftn}


 Before considering quantum theta function,
 we first review the algebra valued inner product on a bimodule
after Rieffel \cite{rief88}.
Let $M$ be any locally compact Abelian group, and $\widehat{M}$ be
its dual group, and let ${\cal G} \equiv M \times \widehat{M} $.
Let $\pi$ be a representation of ${\cal G}$ on $L^2(M)$ such that
\begin{align}
\pi_x \pi_y = \alpha (x,y) \pi_{x+y} =\alpha (x,y)
\overline{\alpha}(y,x) \pi_y \pi_x ~~~ {\rm for}~~ x,y \in {\cal
G} \label{ccl}
\end{align}
where $\alpha$ is a map $ \alpha : ~ {\cal G} \times {\cal G}
\rightarrow {\C}^* $ satisfying
\[ \alpha(x,y)
=\alpha(y,x)^{-1} , ~~~ \alpha(x_1 + x_2 , y) = \alpha(x_1 , y)
\alpha (x_2 , y) ,  \] and $\overline{\alpha}$ denotes the complex
conjugation of $\alpha$.
%
 Let $D$ be a
discrete subgroup of $\cal{G}$. We define $\mathcal{S}(D)$ as the
space of Schwartz functions on $D$.
 For $\Psi \in \mathcal{S}(D)$, it can be expressed as $\Psi = \sum_{w \in D} \Psi(w) e_{D,
 \alpha}(w)$ where $e_{D, \alpha}(w)$ is a delta function with
 support at $w$ and obeys the following relation.
\begin{equation}
e_{D, \alpha} (w_1) e_{D, \alpha} (w_2) = \alpha(w_1,w_2) e_{D,
\alpha} (w_1 +w_2) \label{ccld}
\end{equation}
%



 For Schwartz functions $f,g \in \mathcal{S}(M)$, the algebra ($\mathcal{S}(D)$) valued
inner product is defined as
\begin{align}
{}_D <f,g> \equiv \sum_{w\in D} {}_D<f,g>(w) ~ e_{D, \alpha}(w) ~
\label{aip}
\end{align}
where
\begin{align}
{}_D<f,g>(w) = <f, \pi_w g> . \nonumber
\end{align}
Here, the scalar product of the type $<f,p>$ above with $f,p \in
L^2 (M)$ denotes the following.
\begin{align}
<f,p> = \int f(x_1) \overline{p(x_1)} d \mu_{x_1}  ~~~{\rm for} ~~
x=(x_1,x_2) \in M \times \widehat{M} , \label{sp}
\end{align}
where $\mu_{x_1}$ represents the Haar measure on $M$ and
$\overline{p(x_1)}$ denotes the complex conjugation of $p(x_1)$.
 The $\mathcal{S}(D)$-valued inner product can be represented as
\begin{align}
{}_D <f,g> =\sum_{w\in D} <f, \pi_w g> ~ e_{D, \alpha}(w) ~.
\label{aipr}
\end{align}
%
 For $\Psi \in \mathcal{S}(D) $ and $f \in
\mathcal{S}(M)$, then $~ \pi (\Psi) f \in \mathcal{S}(M)$ can be
written as \cite{rief88}
\begin{align}
(\pi(\Psi)f)(m) & = \sum_{w \in D} \Psi (w) (\pi_w f) (m)
\end{align}
where $m\in M, ~ w \in D \subset M \times \widehat{M}$.


Now, we consider Manin's quantum theta function $\Theta_D$
\cite{manin1-tr,manin2,manin3} for the embedding into vector
space. In \cite{manin3}, quantum theta function was defined via
algebra valued inner product up to a constant factor \cite{ek1},
\begin{align}
{}_D<f , f > &  \sim  \Theta_D , \label{qtheta-def}
\end{align}
where $f$ used in the Manin's construction \cite{manin3} was a
simple Gaussian theta vector
\begin{align}
f = e^{\pi i x_1^t T x_1}, ~~ x_1 \in M. \label{tv-gauss}
\end{align}
Here $T$ is a complex structure given by a complex  skew symmetric
matrix.
%
%
With a given complex structure $T$, a complex variable
$\underline{x} \in {\C}^p$  can be introduced via
\begin{align}
\underline{x} \equiv T x_1 +x_2
\end{align}
where $x=(x_1, x_2) \in M \times \widehat{M}$.

 Based on the defining
concept for quantum theta function (\ref{qtheta-def}), one can
define the quantum theta function $\Theta_D$ in the noncommutative
${\T}^{2p}$ case as
\begin{align}
{}_D<f , f > & = \frac{1}{\sqrt{2^p \det ({\rm Im} ~ T )}}
\Theta_D \label{qtfM}
\end{align}
where $f$ is given by (\ref{tv-gauss}) and $T$ corresponds to
$\Omega$ in (\ref{thevec2}).
According to (\ref{aip}), the $\mathcal{S}(D)$-valued inner
product  (\ref{qtfM}) can be written as
\begin{align}
{}_D<f , f >  =\sum _{h \in D} <f , \pi_h f > e_{D, \alpha} (h) .
\label{sdip}
\end{align}

  In \cite{manin3}, Manin  showed that the quantum theta
function defined in (\ref{qtfM}) is given by
\begin{align}
 \Theta_D & = \sum _{h \in D}  e^{- \frac{\pi}{2} H(\underline{h},\underline{h}) }
  e_{D, \alpha} (h) ,
\label{TDM}
\end{align}
where
\[
H( \underline{g}, \underline{h} ) \equiv \underline{g}^t ( {\rm
Im} T)^{-1} \underline{h}^*
\]
with $ \underline{h}^* = \overline{T} h_1 + h_2 $ denoting the
complex conjugate of $\underline{h}$. At the same time, it also
satisfies a quantum version of the translation action for
classical theta functions \cite{manin1-tr}:
\begin{equation}
{}^\forall g \in D, ~~ C_g ~ e_{D, \alpha} (g) ~ x_g^* ( \Theta_D)
= \Theta_D \label{TDfnr}
\end{equation}
where $C_g$ is defined by
\[ C_g = e^{- \frac{\pi}{2} H(\underline{g},\underline{g})} \]
and the action of $x_g^*$, ``quantum translation", is given by
\begin{align}
x_g^* (e_{D, \alpha} (h)) = e^{- \pi
H(\underline{g},\underline{h})} e_{D, \alpha} (h). \label{xtrans}
\end{align}
In \cite{manin1-tr}, Manin has also required that the factor $
C_g, ~ g \in D $ appearing in the quantum translation $ x_g^* $
has to satisfy the following relation under a combination of
quantum translations for consistency.
\begin{align}
\frac{C_{g+ h}}{C_g C_h} =  {\cal T}_g(h) \alpha(g,h).
\label{xtr-cond}
\end{align}
Here $\alpha(g,h)$ is the cocycle appearing in (\ref{ccld}), and
${\cal T}_g(h)$ is a generalized expression of the factor that
appears by quantum translation:
\begin{align}
x_g^* (e_{D, \alpha} (h)) \equiv  {\cal T}_g(h) e_{D, \alpha} (h).
\label{xtrans-gen}
\end{align}
 The proof of the functional relation (\ref{TDfnr}) in this embedding case
 with quantum translation (\ref{xtrans})
  was shown in \cite{manin3}.


 We now construct the quantum theta function for general
 embedding of ${\R}^p \times {\Z}^q $
 for $2p+q= d$, using the function obtained in the previous
 section.
 With the function
 $f(\vec{s}, \vec{n})$  given by (\ref{thevec2}) we evaluate the quantum
theta function {\it a la} Manin.
\begin{align}
\frac{1}{\sqrt{2^p \det ({\rm Im} \Omega ) }} \hat{\Theta}_{D} =
{}_D<f , f
> , \label{qtfM2}
\end{align}
where $\Omega$ is  a ``complex structure" over the continuous part
of the embedding space as it is determined in the previous section
including the noncommutativity parameters.
We will see that the quantum theta function obtained this way also
satisfies the Manin type functional relation with modified quantum
translation :
\begin{equation}
{}^\forall g \in D, ~~ \hat{C}_g ~ e_{D, \alpha} (g) ~ \hat{x}_g^*
( \hat{\Theta}_{D}) = \hat{\Theta}_D
\label{TDfnr2}
\end{equation}
where $\hat{C}_g, ~ \hat{x}_g^* $ are to be defined below.

To evaluate the quantum theta function (\ref{qtfM2}), we calculate
the scalar product inside the summation in (\ref{sdip}) first.
 For that we first write the action of the operator $\pi_h $ on $f$
 omitting the arrow which denotes a vector for brevity:
\begin{equation}
\pi_h f(s,n)   =  e^{ 2 \pi i ({w_h}_2 \cdot s + r \cdot n) + \pi
i ( {w_h}_1 \cdot {w_h}_2  + m \cdot r )  } f (s+{w_h}_1 , n + m)
, \label{actionforpi}
\end{equation}
where $h \in D$ is given by
\[ h = ( {w_h}_1, {w_h}_2, m, r ) \in {\R}^p \times
{\R}^{p*}  \times {\Z}^q \times {\T}^q  . \]
Then,
\begin{eqnarray}
<f , \pi_h f > & = & \sum_{n \in {\Z}^q} \int_{{\R}^p} ds ~ e^{
\pi [ i s^t  \Omega  s - \frac{1}{2} \sum_{i=1}^{\frac{q}{2}}
(n_i^2 +n_{\frac{q}{2} +i}^2)]}
  e^{\pi [-2i ({w_h}_2 \cdot s + r \cdot n) - i ( {w_h}_1
\cdot {w_h}_2  + m \cdot r )]  }
\nonumber \\
&  & ~~~ \hspace*{1cm} \times e^{  \pi [ - i (s+ {w_h}_1)^t \Omega
(s+ {w_h}_1) -  \frac{1}{2} \sum_{i=1}^{\frac{q}{2}} ((n_i +
m_i)^2 +(n_{\frac{q}{2} +i} +m_{\frac{q}{2} +i})^2)]}
\nonumber \\
& = &  \int_{{\R}^p} d s ~ e^{ -2\pi [ s^t  ({\rm Im} \Omega) s +
i {w_h}_1^t  \overline{\Omega}  s + i {w_h}_2 \cdot s] - i \pi
[{w_h}_1^t  \overline{\Omega}  {w_h}_1 + {w_h}_1 \cdot {w_h}_2 ]}
\nonumber \\
&  &  \times  e^{ - \frac{\pi}{2} \sum_{i=1}^{\frac{q}{2}} (m_i^2
+m_{\frac{q}{2} +i}^2)  - \pi i  m \cdot r  } \sum_{n \in {\Z}^q}
 e^{- \frac{\pi}{2} \sum_{i=1}^{\frac{q}{2}} (n_i^2
+n_{\frac{q}{2} +i}^2) +  2 \pi i [ n \cdot ( -r +  \frac{i m}{2})
]}
 \nonumber \\
& = & \prod_{j=1}^{q} b_{r_j, m_j}  \int_{{\R}^p} d s ~ e^{ -2\pi
[ s^t  ({\rm Im} \Omega) s + i {w_h}_1^t  \overline{\Omega}  s + i
{w_h}_2 \cdot s] - i \pi [{w_h}_1^t  \overline{\Omega} {w_h}_1 +
{w_h}_1 \cdot {w_h}_2 ]} , \label{scprod}
\end{eqnarray}
where
\begin{equation}
 b_{r_j, m_j} =  e^{- \frac{\pi}{2} m_j^2 - \pi i m_j r_j}
                ~ \theta (\tau =  i, ~ z= -r_j + \frac{im_j}{2}),
  ~~ j=1, \cdots, q .
\label{c-latt}
\end{equation}
Here, $\theta (\tau , z)$ is the classical theta function defined
by
\[
\theta (\tau ,  z) = \sum_{n \in {\Z}} e^{ \pi i  \tau n^2 +  2
\pi i n z }, ~ ~ {\rm for} ~ ~ \tau , z \in {\C} .
\]
The integral in (\ref{scprod}) is the same as that appeared in
\cite{manin3} and is given by
\begin{equation}
\label{contint} \frac{1}{\sqrt{2^p \det ({\rm Im} \Omega )}} e^{-
\frac{\pi}{2} H(\underline{w_h}, \underline{w_h})}  .
\end{equation}
Thus we obtain the following result.
\\
\begin{proposition}
The quantum theta function $\hat{\Theta}_D$ obtained from $f$ in
\eqref{thevec2} is given by
\begin{align}
 \hat{\Theta}_D & = \sum _{h \in D} \widetilde{b}_h ~ e^{- \frac{\pi}{2} H(\underline{w_h},\underline{w_h}) }
  e_{D, \alpha} (h) ,
\label{TDM2}
\end{align}
where
\begin{equation}
\widetilde{b}_h = \prod_{j=1}^q b_{r_j, m_j} \label{lat-factor}
\end{equation}
with $b_{r_j, m_j}$ given in \eqref{c-latt}.
\end{proposition}
The above quantum theta function satisfy the Manin's functional
relation under ``modified quantum translation" (\ref{TDfnr2}), and
we get the following theorem.
\\
\begin{theorem}
\begin{equation*}
{}^\forall g \in D, ~~ \hat{C}_g ~ e_{D, \alpha} (g) ~ \hat{x}_g^*
( \hat{\Theta}_{D})
 = \hat{\Theta}_{D} ,
\end{equation*}
and the consistency condition (\ref{xtr-cond}) for $\hat{C}_g$.
The above relation is satisfied if we assign
\begin{equation}
\hat{C}_g = \widetilde{b}_g ~ e^{- \frac{\pi}{2}
H(\underline{w_g},\underline{w_g})} ,
\label{trc-lat}
\end{equation}
and $ \hat{x}_g^* $ is defined by
\begin{equation}
\hat{x}_g^* (e_{D, \alpha} (h)) = \hat{{\cal T}}_g(h) e_{D,
\alpha} (h) \label{tr-latt}
\end{equation}
with
\begin{align}
\hat{{\cal T}}_g(h) = \frac{\hat{C}_{g+ h}}{\hat{C}_g \hat{C}_h
\alpha(g,h) } .
\label{xtr2-cond}
\end{align}
\end{theorem}
\textbf{Proof.} Now, it is easy to show the relation (\ref{TDfnr2}):
\begin{eqnarray*}
 \hat{C}_g ~ e_{D, \alpha} (g) ~ \hat{x}_g^* ( \hat{\Theta}_{D})
& = & \hat{C}_g ~ e_{D, \alpha} (g) ~ \hat{x}_g^* (\sum _{h \in D}
\widetilde{b}_h ~ e^{- \frac{\pi}{2}
H(\underline{w_h},\underline{w_h}) }
  e_{D, \alpha} (h)) \\
& = & \hat{C}_g ~ e_{D, \alpha} (g) ~ \hat{x}_g^* (\sum _{h \in D}
\hat{C}_h e_{D, \alpha} (h))\\
& = & \sum _{h \in D} \hat{C}_g \hat{C}_h e_{D, \alpha} (g)
\hat{{\cal T}}_g(h) e_{D, \alpha} (h)\\
 & = &  \sum _{h \in D} \hat{C}_{g+ h} e_{D, \alpha} (g+h)   = \hat{\Theta}_{D} .
\end{eqnarray*}
where we used the relation (\ref{trc-lat}) in the second step, and
the relation (\ref{xtr2-cond}) together with the cocycle condition (\ref{ccld})
in the last step.
\\

\noindent
\textbf{Remark.} Here we notice that the quantum
translations are not additive in this case:
\begin{align}
\hat{x}_{g_1}^* \cdot \hat{x}_{g_2}^* (e_{D, \alpha} (h)) \neq
\hat{x}_{g_1 + g_2}^* (e_{D, \alpha} (h)) . \label{qtr-latt}
\end{align}
On the other hand, the quantum translations in the Manin's case
($x_g^*$), (\ref{xtrans}), are additive:
\begin{align}
x_{g_1}^* \cdot x_{g_2}^* (e_{D, \alpha} (h)) = x_{g_1 + g_2}^*
(e_{D, \alpha} (h)) . \label{qtr-vectsp}
\end{align}
\\
%
%




\section*{4. Conclusion }

In this paper, we  study the theta vector and the corresponding
quantum theta function for noncommutative tori with general
embeddings.

While the theta vector exists in the embedding into vector space
case ($ {\R}^p $ type), there does not exist fully holomorphic
theta vector in the embedding into lattice case (${\Z}^q $ type).
We construct a module which consists of holomorphic vectors for
the vector space part and a plain Schwartz function for the
lattice part in the case of mixed embedding ($ {\R}^p \times
{\Z}^q $ type). Manin has constructed the quantum theta functions
only with holomorphic modules with embedding into vector space.
And, it was not clear whether the partially holomorphic modules
such as ours for mixed embeddings would yield the quantum theta
functions that satisfy the Manin's requirement.
The answer turns out to be yes.

There is one differenence between the two types of quantum theta
functions, Manin's and ours. In the Manin's quantum theta
function, two consecutive ``quantum translations" are additive,
while those of ours are not. This non-additivity is allowed by the
consistency condition for the cocycle and quantum translation,
(\ref{xtr2-cond}).

In conclusion, we have shown that the quantum theta functions on
noncommutative tori that satisfy the Manin's requirement can be
constructed with any choice of the following embeddings, 1) into
vector space times lattice, 2) into vector space, 3) into lattice.
Our result for the cases 1) and 3) can be directly extended to the
embeddings that include finite groups as was done in the Manin's
work \cite{manin3} for the case 2).
\\

\vspace{5mm}

\noindent
{\Large \bf Acknowledgments}

\vspace{5mm} \noindent The authors thank KIAS for hospitality
during the time that this work was done. This work was supported
by Korean Council for University Education, grant funded by Korean
Government(MOEHRD) for 2006 Domestic Faculty Exchange (E. C.-Y.),
and by KOSEF Research Grant No. R01-2006-000-10638-0 (H. K.).
\\




\end{document}